\documentclass[useAMS,usenatbib]{mn2e}
\usepackage{multirow}
\usepackage{epsfig}
\usepackage{subfigure}
\usepackage{url}

\makeatletter
\let\jnl@style=\rm
\def\ref@jnl#1{{\jnl@style#1}}

\def\aj{\ref@jnl{AJ}}                   
\def\araa{\ref@jnl{ARA\&A}}             
\def\apj{\ref@jnl{ApJ}}                 
\def\apjl{\ref@jnl{ApJ}}                
\def\apjs{\ref@jnl{ApJS}}               
\def\ao{\ref@jnl{Appl.~Opt.}}           
\def\apss{\ref@jnl{Ap\&SS}}             
\def\aap{\ref@jnl{A\&A}}                
\def\aapr{\ref@jnl{A\&A~Rev.}}          
\def\aaps{\ref@jnl{A\&AS}}              
\def\azh{\ref@jnl{AZh}}                 
\def\baas{\ref@jnl{BAAS}}               
\def\jrasc{\ref@jnl{JRASC}}             
\def\memras{\ref@jnl{MmRAS}}            
\def\mnras{\ref@jnl{MNRAS}}             
\def\pra{\ref@jnl{Phys.~Rev.~A}}        
\def\prb{\ref@jnl{Phys.~Rev.~B}}        
\def\prc{\ref@jnl{Phys.~Rev.~C}}        
\def\prd{\ref@jnl{Phys.~Rev.~D}}        
\def\pre{\ref@jnl{Phys.~Rev.~E}}        
\def\prl{\ref@jnl{Phys.~Rev.~Lett.}}    
\def\pasp{\ref@jnl{PASP}}               
\def\pasj{\ref@jnl{PASJ}}               
\def\qjras{\ref@jnl{QJRAS}}             
\def\skytel{\ref@jnl{S\&T}}             
\def\solphys{\ref@jnl{Sol.~Phys.}}      
\def\sovast{\ref@jnl{Soviet~Ast.}}      
\def\ssr{\ref@jnl{Space~Sci.~Rev.}}     
\def\zap{\ref@jnl{ZAp}}                 
\def\nat{\ref@jnl{Nature}}              
\def\iaucirc{\ref@jnl{IAU~Circ.}}       
\def\aplett{\ref@jnl{Astrophys.~Lett.}} 
\def\apspr{\ref@jnl{Astrophys.~Space~Phys.~Res.}}
\def\bain{\ref@jnl{Bull.~Astron.~Inst.~Netherlands}}
\def\fcp{\ref@jnl{Fund.~Cosmic~Phys.}}  
\def\gca{\ref@jnl{Geochim.~Cosmochim.~Acta}}   
\def\grl{\ref@jnl{Geophys.~Res.~Lett.}} 
\def\jcp{\ref@jnl{J.~Chem.~Phys.}}      
\def\jgr{\ref@jnl{J.~Geophys.~Res.}}    
\def\jqsrt{\ref@jnl{J.~Quant.~Spec.~Radiat.~Transf.}}
\def\memsai{\ref@jnl{Mem.~Soc.~Astron.~Italiana}}
\def\nphysa{\ref@jnl{Nucl.~Phys.~A}}   
\def\physrep{\ref@jnl{Phys.~Rep.}}   
\def\physscr{\ref@jnl{Phys.~Scr}}   
\def\planss{\ref@jnl{Planet.~Space~Sci.}}   
\def\procspie{\ref@jnl{Proc.~SPIE}}   

\makeatother

\title[X-ray absorption variability in NGC 4507]{X-ray absorption variability in NGC 4507}

\author[Andrea Marinucci, et al.]{Andrea Marinucci$^{1,2}$\thanks{E-mail: marinucci@fis.uniroma3.it (AM)},  Guido Risaliti$^{2,3}$, Junfeng Wang$^2$ , Stefano Bianchi$^{1}$,  \newauthor   Martin Elvis$^2$, Giorgio Matt$^1$, Emanuele Nardini$^2$, Valentina Braito$^3$ \\
$^1$Dipartimento di Fisica, Universit\`a degli Studi Roma Tre, via della Vasca Navale 84, 00146 Roma, Italy\\
$^2$Harvard-Smithsonian Center for Astrophysics, 60 Garden St., Cambridge MA 02138, USA\\
$^3$INAF - Osservatorio Astrofisico di Arcetri, L.go E. Fermi 5, Firenze, Italy\\
$^4$INAF - Osservatorio Astronomico di Brera, Via E. Bianchi 46, I-23807, Merate, Italy\\
}

\begin{document}
\maketitle
\label{firstpage}

\begin{abstract} 
We present a complete spectral analysis of an XMM-Newton and Chandra campaign of the obscured AGN in NGC 4507, consisting of six observations spanning a period of six months, ranging from June 2010 to December 2010.
We detect strong absorption variability on time scales between 1.5 and 4 months, suggesting that the obscuring material consists of gas clouds at parsec-scale distance. 
The lack of significant variability on shorter time scales suggests that this event is not due to absorption by broad line region clouds, which was instead found in other studies of similar sources. This shows that a single, universal structure of the absorber (either BLR clouds, or the parsec-scale torus) is not enough to reproduce the observed complexity of the X-ray absorption features of this AGN.
\end{abstract}


\section{ Introduction}
{\bf The standard Unification Model} for active galactic nuclei (AGN) assumes the same internal structure for Seyfert 2 and Seyfert 1 galaxies \citep{antonucci93}, with all the type 1/2 observational differences ascribed to an axisymmetric absorber/reflector, located between the broad line region and the narrow line region, in order to obscure the former, but not the latter.  An early developed, natural geometrical and physical scenario is that of a homogeneous torus on a parsec scale \citep{krbe88}; however recent studies on  X-ray absorbing column density changes performed with {\itshape Chandra, XMM-Newton} and {\itshape Suzaku} satellites ruled out a universal geometrical structure of the circumnuclear absorber. Absorption variability is common (almost ubiquitous) when we compare observations months to years apart \citep{risa02b}, and, most notably, has been found on time scales of hours to days in several sources, such as NGC1365 \citep{ris05, ris07, ris09}, NGC 4388 \citep{elvis04}, NGC~4151 \citep{puc07} and NGC 7582 \citep{bianchi09c}. 

NGC 4507 is a nearby (z=0.0118) barred spiral galaxy and one of the X-ray brightest Compton-thin Seyfert 2s, despite the heavy obscuration ($N_{\rm H}\sim 4-9 \times 10^{23}$cm$^{-2}$). 
 It was first observed in the X-rays by Einstein \citep[L$_{\rm X}=2.8\times 10^{41}$ erg s$^{-1}$,][]{kriss80} and  then in 1990 with Ginga \citep{awaki91b}, showing a strongly absorbed ($N_{\rm H}\sim 5 \times 10^{23}$cm$^{-2}$) power law continuum and a prominent iron K$\alpha$ line. In 1994 ASCA also revealed a strong X-ray excess and an intense emission line (identified as Ne \textsc{ix}) at $\sim 0.9$ keV \citep{comastri98}. \citet{risa02} reported then three BeppoSAX observations of the source, confirming the obscuring column densities observed in the previous two observations. In 2001 NGC 4507 has been observed for the first time with the most sensitive X-ray satellites, XMM-Newton and Chandra/ACIS-S HETG \citep{matt04b} confirming once again the clear Compton-thin state of the spectrum, with a $\Gamma=1.8_{-0.2}^{+0.1}$ power law absorbed by an $N_{\rm H}=4.4_{-0.6}^{+0.5}\times 10^{23}$cm$^{-2}$.
In 2007 a $\sim$3 days observation with  Suzaku has been performed and it revealed a much larger absorbing column density ($N_{\rm H} \sim9\times 10^{23}$cm$^{-2}$) with respect to the earlier observations, but no
changes {\em within} the 3 days of monitoring \citep[][hereafter B12]{braito12}.
Summarizing, NGC 4507 showed strong variations in time scales of years (from 1990 until 2007) but none in shorter time scales (3 days observation with Suzaku). This excluded a possible sub-parsec scale absorbing structure, but leaves the actual size and distance of the absorber largely unconstrained.
In this paper we report the study of five XMM-Newton observations spanning a period of 6 weeks, between June and August~2010, and a Chandra observation performed 4 months later, in December 2010. This observational campaign has been designed to fill the time gap between days and years in the previous observations, so constraining the location of the absorber. 

\section{Observations and data reduction}
The 5 XMM-\textit{Newton} observations analysed in this paper were performed on 2010 June 24 (obsid 0653870201), July 3 (obsid 0653870301), July 13 (obsid 0653870401), July 23 (obsid 0653870501), August 3 (obsid 0653870601) with the EPIC CCD cameras, 
the PN \citep{struder01} and the two MOS \citep{turner01}, operated in large window and medium filter mode. The extraction radii and the optimal time cuts for flaring particle background were computed with SAS 11 \citep{gabr04} via an iterative process which leads to a maximization of the Signal-to-Noise Ratio (SNR), similarly to that described in \citet{pico04}. After this process, the net exposure times for the 5 different observations were 16 ks, 13 ks, 13 ks, 13 ks and 17 ks for the PN, respectively. The resulting optimal extraction radii are 40 arcsec for the first three observations, 30 arcsec for the fourth one, 26 arcsec for the last one. The background spectra were extracted from source-free circular regions with a radius of about 50 arcsec for all the 5 observations. We also re-extracted the data from a previous XMM-\textit{Newton} observation (obsid 0006220201), with a net exposure of about 36 ks, adopting an extraction radius of 40 arcsec for the source and 42 arcsec for the background. The analysis of the last set of data is discussed in \citet{matt04b}. \\
\textit{Chandra} observed the source on December~2, 2010 for 44~ks, with the Advanced
CCD Imaging Spectrometer \citep[ACIS:][]{acis}. Data were reduced with the Chandra Interactive Analysis
of Observations \citep[CIAO:][]{ciao} 4.4 and the
Chandra Calibration Data Base (CALDB) 4.4.6 database,
adopting standard procedures, using a 2 arcsec and 10 arcsec extraction radii for the source and background, respectively. \\
Spectra were binned in order to over-sample the instrumental resolution by at least a factor of 3 and to have no less than 30~counts in each background-subtracted spectral channel. This allows the applicability of the $\chi^2$ statistics.

\section{Data analysis}
The adopted cosmological parameters are $H_0=70$ km s$^{-1}$ Mpc$^{-1}$, $\Omega_\Lambda=0.73$ and $\Omega_m=0.27$, i.e. the default ones in \textsc{xspec 12.7.0} \citep{xspec}. Errors correspond to the 90\% confidence level for one interesting parameter ($\Delta\chi^2=2.7$), if not otherwise stated. 

\subsection{\label{EPICanalysis}EPIC PN/MOS spectral analysis}
The soft 0.5-3.0 keV spectrum presents a strong 'soft excess', which appears dominated by emission lines from an highly ionized gas, as observed in most X-ray obscured AGN \citep{turner97, guabia07}. Emission lines from H-like and He-like C, N, O, and Ne, as well as from the Fe L-shell, have been detected and reported in previous observations \citep{matt04b}.\\
Following \citet{matt04b} the baseline model we used to fit the 0.5-10 keV spectra can be roughly expressed by the following general formula:
\begin{eqnarray} 
 F(E)= e^{-\sigma(E) N_{\rm H}^G}[Ph_C + C + e^{-\sigma(E) N_{\rm H}}BE^{-\Gamma} + \\ \nonumber
 + R(\Gamma)+ \sum_i G_i(E)] \ \ \ \ \ \ \ \ \ \ \ \ \ \ \ \ \ \ \ \ \ \ \ \ \ \ \ \ \ \ 
\end{eqnarray}
where $\sigma(E)$ is the photoelectric cross-section \citep[abundances as in][]{angr89}, $N_{\rm H}^G$ is the Galactic absorbing column density along the line of sight to the source \citep{dl90}; $Ph_C$ is the emission from a photoionised gas reproduced with self-consistent \textsc{cloudy} models as described in \citet{bianchi10} and \citet{marinucci11} while C is the emission from a collisionally-ionised diffuse gas \citep[\textsc{apec} model,][]{apec}; $N_{\rm H}$ is the neutral absorbing column density at the redshift of the source; B is the normalization of the primary powerlaw with slope $\Gamma$; $R(\Gamma)$ is the Compton scattering from the inner layer of the circumnuclear torus, modelled in \textsc{Xspec} with {\sc pexrav} \citep{mz95};$G_i(E)$ are Gaussian profiles, corresponding to required emission lines of high-Z elements such as the Fe K$\alpha $ at 6.4 keV, Fe K$\beta$ at 7.058 keV, Fe \textsc{xxvi} at 6.966 keV and the forbidden line of the Fe \textsc{xxv} K$\alpha$ triplet (see table \ref{bestfit}).\\
A further Gaussian emission line has been used to reproduce the Compton Shoulder (CS) redwards of the Iron line core, as expected on theoretical grounds, with energy fixed at 6.3 keV and $\sigma$=40 eV \citep{matt02}.\\
\begin{figure}
\begin{center}
\epsfig{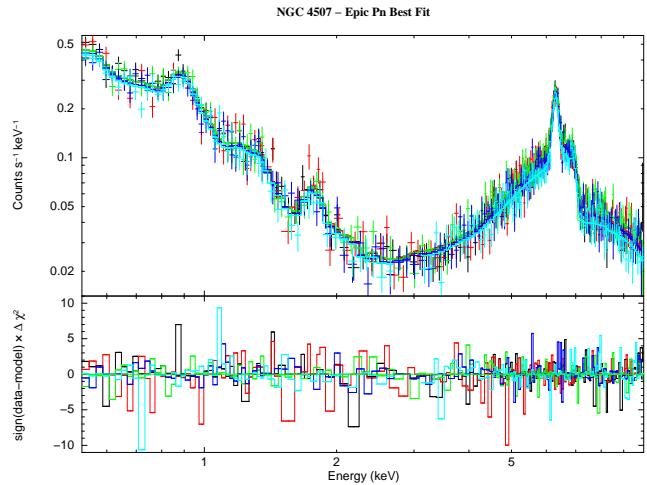}
\caption{\label{fig1} XMM-Newton EPIC PN 0.5-10 keV best fit and residuals for the 5 observations of the campaign described in this work.}
\end{center}
\end{figure}
The reflected, Compton scattered emission has been modeled using the \textsc{pexrav} model with $\Gamma$=1.8 and normalization fixed to the values measured with the broadband Suzaku observation  (B12).
The previous model has been used to fit, in first place, the 5 separate EPIC PN/MOS observations. 
The 5 spectral fits from our campaign in Summer 2010 (labeled as Obs. 1-5 in Table \ref{bestfit}) have good $\chi^2/$d.o.f. and do not present any strong evidence of variations in either the absorbing column density or $\Gamma$. The Obs. 2 data set presents strong residuals mainly between 1.5 and 3.0 keV, more evident in the EPIC-MOS spectra. Since they are located in very narrow bins at $\sim2.8$, $\sim1.6$ and $\sim1.9$ keV and they cannot be ascribed to known spectral features we believe  they are due to background/calibration issues.

Two different photoionised phases and a collisional one are needed to model the 0.5-3.0 keV spectra of NGC 4507. Table \ref{bestfit} clearly shows that the soft X-ray emitting gas has not varied during the monitoring. This suggests that the interveining absorbing material, responsible for the column density variation between June and December 2010, might be much closer to the X-ray source than the circumnuclear matter responsible for the soft emission. As already discussed in \citet{bianchi06} and \citet{bg07} this gas is likely coincident with the NLR. Further studies on the phoionisation mechanisms and extended emission in NGC 4507 will be discussed in detail in the future (Wang et al., in preparation).

We then analysed the 10 spectra (5 EPIC PN and 5 MOS1+2, labeled as Set 1 in Table 1) simultaneously, using the model described above and linking all the parameters, except for the flux normalizations. The  baseline model reproduces the 5 sets of data and some residuals are present around 1.8 keV. Indeed, the addition of a line at 1.77$\pm0.03$ keV is required  ($\Delta \chi^2=44$, with a significance greater than 99.99\%, according to {\it F}-test\footnote{ The F-test is not a reliable test for the significance of emission lines unless their normalizations are allowed to be negative \citep{prota02}.}); it can be identified as Si K$\alpha$, with a corresponding flux of $2.0\pm0.4\times 10^{-6}$ ph cm$^{-2}$ s$^{-1}$ . The primary powerlaw ($\Gamma=1.8^{+0.2}_{-0.2}$) is absorbed by a column density of $9.0\pm0.5 \times 10^{23} $cm$^{-2}$. The photon index is in agreement with previous studies on this source, on the contrary a clear variation in the N$_{\rm H}$ can be noticed with respect to the old 2004 XMM-Newton observation, but it is fully compatible with the Suzaku observation in 2007 (Fig. \ref{nhplot}).
The addition of a CS redwards of the Iron K$\alpha$ line core is required by the fit ($\Delta \chi^2=20$ with a significance greater than 99.99\%) and its flux, being $24\pm7\%$ of the flux of the narrow core of the Fe K$\alpha$, is consistent with expectations. Both Iron K$\alpha$ and CS fluxes are in agreement with the ones found in the previous XMM observation  \citep{matt04b}.
The equivalent widths of the high energy emission lines are in full agreement with the ones we find when we analyse the 5 sets of data separately. We only find upper limits on fluxes ($<0.5\times 10^{-6}$ ph cm$^{-2}$ s$^{-1}$) and EWs ($<50$ eV) of the Fe \textsc{xxvi} emission lines at 6.966 keV.

As a last cross-check, we let the 5 different absorbing column density parameters free to vary in our fit, to check whether a possible variation may have been occurred in the 1.5 months monitoring. The best fit values of the 5 different N$_{\rm H}$ do not show any significant variation with respect to the best fit value for the whole data set ($9.0\pm0.5\times 10^{23} $ cm$^{-2}$) with a non significant  improvement of the fit ($\Delta \chi^2=5$ with four more parameters and a significance lower than 8\%). This result brings further evidence to the argument that absorption variability on short time scales (hours-days) can be ruled out in our analysis of NGC 4507. If the intervening absorbing material had varied on such short time scales we would not have measured a constant column density in a 1.5 months monitoring. The measured values would have been completely scattered over the range of values observed in the past ($4-9 \times 10^{23} $cm$^{-2}$).

\begin{figure}
\begin{center}
\epsfig{file=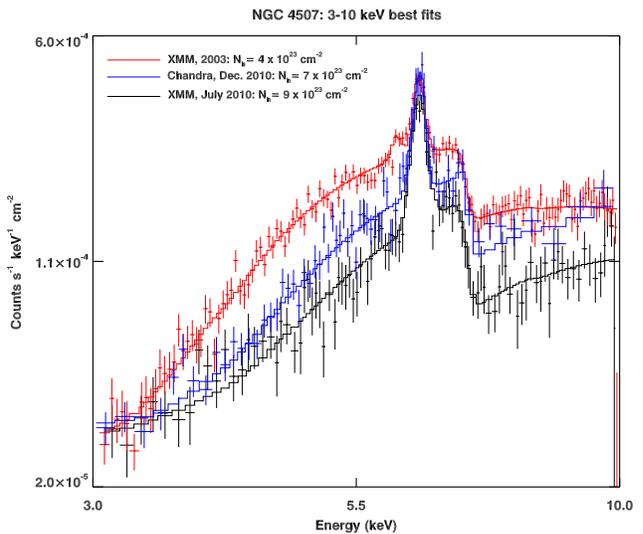, width=\columnwidth}
\caption{\label{310keV} 3-10 keV best fits: each spectrum is divided by the effective area of the instrument. The impact of the variation in the absorbing column density on the three spectra can be clearly seen. }
\end{center}
\end{figure}

\subsection{\label{chandraanalysis}Chandra ACIS spectral analysis}
The baseline model is the same we used to fit the XMM-Newton data. The overall fit is very good ($\chi^2= $327/329) and the addition of a further emission line at 1.81$\pm0.02$ keV is required ($\Delta \chi^2=14$ with a significance greater than 99.99\%), with a flux of $4 {+1 \atop -2} \times 10^{-6} \,\rm ph \, cm^{-2} \, s^{-1}$, marginally consistent with the emission from Si K$\alpha$ already found in XMM-Newton best fits. The soft X-ray spectrum is produced by two photoionised phases, while the contribution by a collisional gas is not required by the fit. Equivalent widths, fluxes and energy centroids of the emission lines found in the 5 - 7.5 keV energy range are fully consistent with the results reported above. Best fit values are shown in Table \ref{bestfit} and the column is labeled as Obs. 6.\\
The reflected primary continuum has been fitted as described before (see Sect. \ref{EPICanalysis}): the best fit value of the absorbing column density is $6.5\pm0.7\times 10^{23} $ cm$^{-2}$, leading to a $2.5\times 10^{23}$~cm$^{-2}$ variation at a $3\sigma$ confidence level in a time scale ranging between 1.5 (time interval between the first and the fifth XMM-Newton observation) and 4 months (time interval between the last XMM-Newton observation and the one with Chandra). In Fig. \ref{310keV} we show the influence of the change in N$_{\rm H}$ on the spectral shape between 3 and 10 keV. Considering the partial degeneracy between the column density and the spectral slope, the significance of the variation is even stronger, as illustrated by the $\Gamma$-N$_H$ contour plots shown in Fig.~3.

\begin{table*}
\begin{center}
\begin{tabular}{ccccccccc}
& & & & & & & \\
\hline
\hline
{\bfseries Parameter} &   {\bfseries Obs. 1  } &{\bfseries Obs. 2  }  &{\bfseries Obs. 3  } & {\bfseries Obs. 4} & {\bfseries Obs.  5} & {\bfseries Obs.  6} & {\bfseries Set 1}   \\
\hline
K (PN-MOS)  &$0.99^{+0.02}_{-0.02}$ & $1.06^{+0.03}_{-0.03}$ & $1.02^{+0.03}_{-0.03}$ & $1.02^{+0.03}_{-0.03}$ & $0.93^{+0.03}_{-0.03}$&- &- \\
& & & & & & & &\\
$N_{\rm H}$ ($10^{23}$ cm${}^{-2}$) & $8.7^{+0.7}_{-0.8}$& $9.7^{+0.9}_{-0.9}$&$7.6^{+1.0}_{-1.3}$&$9.4^{+1.1}_{-1.1}$  & $8.0^{+0.8}_{-0.6}$& $6.5^{+0.7}_{-0.7}$& $9.0^{+0.5}_{-0.5}$\\
& & & & & &  &\\
$\Gamma$  & $1.8^{+0.1}_{-0.1}$ & $2.0^{+0.3}_{-0.3}$& $1.4^{+0.4}_{-0.3}$&$1.9^{+0.3}_{-0.3}$ &   $1.6^{+0.3}_{-0.4}$& $1.5^{+0.4}_{-0.4}$& $1.8^{+0.2}_{-0.2}$ \\
& & & & & & &  &\\
F (CS${}_{6.3\ \rm keV}$)  & $<1.5$& $<1.7$&$1.8^{+0.7}_{-0.9}$ & $1.4^{+0.7}_{-0.8}$& $0.8^{+0.4}_{-0.4}$&$2.0^{+0.9}_{-1.1}$ & $1.0^{+0.2}_{-0.2}$  \\
 & & & & & & & &\\
Fe K$\alpha$    E&$6.40^{+0.01}_{-0.01}$ & $6.40^{+0.01}_{-0.01}$ & $6.41^{+0.02}_{-0.02}$&  $6.41^{+0.01}_{-0.02}$& $6.40^{+0.02}_{-0.02}$ &  $6.40^{+0.02}_{-0.02}$ &$6.401^{+0.003}_{-0.003}$\\
$\ \ \ \ \ \ \ \ \ \ \ \ \ $ Flux& $4.4^{+0.3}_{-0.4}$& $3.8^{+0.5}_{-0.5}$& $3.7^{+0.7}_{-0.7}$&$4.3^{+0.6}_{-0.7}$ &  $4.6^{+0.5}_{-0.5}$& $4.0^{+0.9}_{-1.0}$& $4.2^{+0.3}_{-0.2}$\\
$\ \ \ \ \ \ \ \ \ \ \ \ $ EW& $315 ^{+15}_{-35}$ &  $250 ^{+40}_{-40}$&$245 ^{+35}_{-45}$ & $285 ^{+40}_{-45}$& $335^{+40}_{-35}$&$220^{+50}_{-40}$ \\
 & & & & & & & &\\
Fe K$\beta$    E& 7.058 &7.058 & 7.058& 7.058&  7.058& 7.058 & 7.058\\
$\ \ \ \ \ \ \ \ \ \ \ \ \ $ Flux&$ <0.6$& $ <0.5$& $ <0.6$& $ <0.5$&  $<0.8$ & $ <1.1$& $0.3^{+0.2}_{-0.2}$\\
$\ \ \ \ \ \ \ \ \ \ \ \ $ EW&$<50$&$<35$ &$<45$ &$<40$ &$<70$ & $<70$\\
& &  & & & & & &\\
Fe \textsc{xxv} E& 6.700& 6.700& 6.700&6.700 & 6.700 &6.700 &6.700  \\
$\ \ \ \ \ \ \ \ \ \ \ \ \ $ Flux&$0.6^{+0.3}_{-0.4}$ &$<0.3$ & $0.7^{+0.4}_{-0.4}$&$<0.7$  &$<0.3$  & $ <0.4$& $0.3^{+0.1}_{-0.1}$\\
$\ \ \ \ \ \ \ \ \ \ \ \ $ EW& $40 ^{+20}_{-25}$&$<20$ &$45 ^{+25}_{-25}$ & $<45$& $<20$& $<20$\\ 
\hline
kT ($\rm keV$)& $0.43^{+0.12}_{-0.08}$& $0.53^{+0.30}_{-0.25}$ &$0.44^{+0.26}_{-0.15}$ & $0.43^{+0.17}_{-0.15}$& $0.42^{+0.12}_{-0.12}$&- & $0.43^{+0.11}_{-0.07}$\\
& &  & & & & & &\\
$\log U_1$ &$1.63^{+0.05}_{-0.05}$ &$1.59^{+0.10}_{-0.07}$  & $1.75^{+0.05}_{-0.10}$& $1.59^{+0.11}_{-0.07}$& $1.67^{+0.07}_{-0.10}$ & $1.85^{+0.20}_{-0.12}$&$1.64^{+0.04}_{-0.04}$ \\
& &  & & & & & &\\
$\log N_{H1}$ & $20.5^{+0.2}_{-0.2}$& $20.9^{+0.3}_{-0.2}$ & $20.9^{+0.2}_{-0.2}$&$21.1^{+0.4}_{-0.3}$& $20.8^{+0.4}_{-0.3}$&$21.9^{+0.8}_{-0.6}$ & $21.0^{+0.1}_{-0.2}$\\
& &  & & & & & &\\
$\log U_2$ & $-0.69^{+0.10}_{-0.25}$& $-0.18^{+0.28}_{-0.23}$ & $-0.45^{+0.30}_{-0.25}$& $-0.47^{+0.64}_{-0.82}$&  $0.30^{+0.30}_{-0.50}$& $0.90^{+0.10}_{-0.40}$&$-0.15^{+0.21}_{-0.25}$\\
& &  & & & & & &\\
$\log N_{H2}$ & $19.9^{+0.1}_{-0.3}$& $20.4^{+0.5}_{-0.5}$ & $<19.8$ &$19.8^{+0.5}_{-0.5}$ & $19.9^{+0.3}_{-0.5}$&$21.5^{+0.1}_{-0.1}$ &$19.9^{+0.2}_{-0.3}$\\
& &  & & & & & &\\
F$_{0.5-2\ \rm keV}$& $0.4^{+0.1}_{-0.1}$&$0.4^{+0.1}_{-0.1}$ &$0.4^{+0.2}_{-0.2}$ & $0.4^{+0.1}_{-0.1}$& $0.4^{+0.2}_{-0.1}$&$0.3^{+0.2}_{-0.2}$ & \\
& &  & & & & & &\\
F$_{2-10\ \rm keV}$& $7.7^{+0.3}_{-0.3}$& $8.0^{+0.3}_{-0.3}$ &$8.4^{+0.2}_{-0.2}$ & $8.0^{+0.3}_{-0.3}$& $7.5^{+0.8}_{-0.6}$&$10.0^{+0.4}_{-0.4}$ & \\
& &  & & & & & &\\
$\chi ^2/$d.o.f.& 308/292& 347/268& 242/272& 267/267& 280/283 &313/327 & 1576/1459\\
\hline
\hline
\end{tabular}\\
\caption{\label{bestfit} Best fit values. Energies are in keV, line fluxes in $10^{-5}$ ph cm$^{-2}$ s$^{-1}$, observed fluxes in  $10^{-12}$ erg cm$^{-2}$ s$^{-1}$ and EWs in eV. Photoionisation parameters $\log U_1$, $\log U_2$ and column densities $\log N_{\rm H1}$, $\log N_{\rm H2}$ are the best fit values of the two photoionised phases needed to reproduce the soft emission; kT is the energy of the additional collisional phase (see text for detail).}
\end{center}
\end{table*}

\begin{figure}
\begin{center}
\epsfig{file=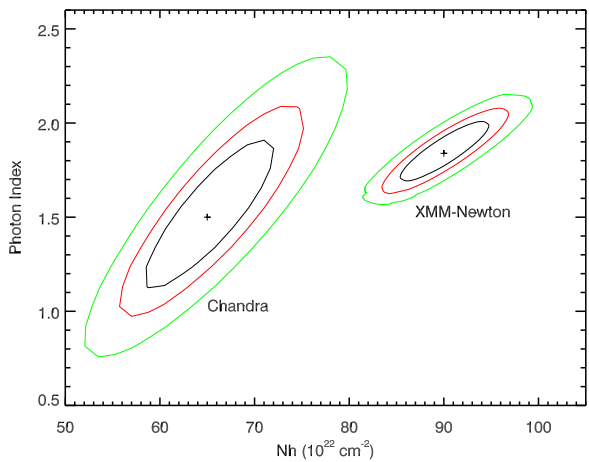,width=\columnwidth}
\caption{\label{contours} $\Gamma$-N$_H$ contour plots of the 5 combined XMM-Newton observations and the {\em Chandra} December~2010 observation. Solid black, red and green lines corresponds to 1$\sigma$, 2$\sigma$, 3$\sigma$ confidence levels, respectively. }
\end{center}
\end{figure}
\begin{figure}
\begin{center}
\epsfig{file=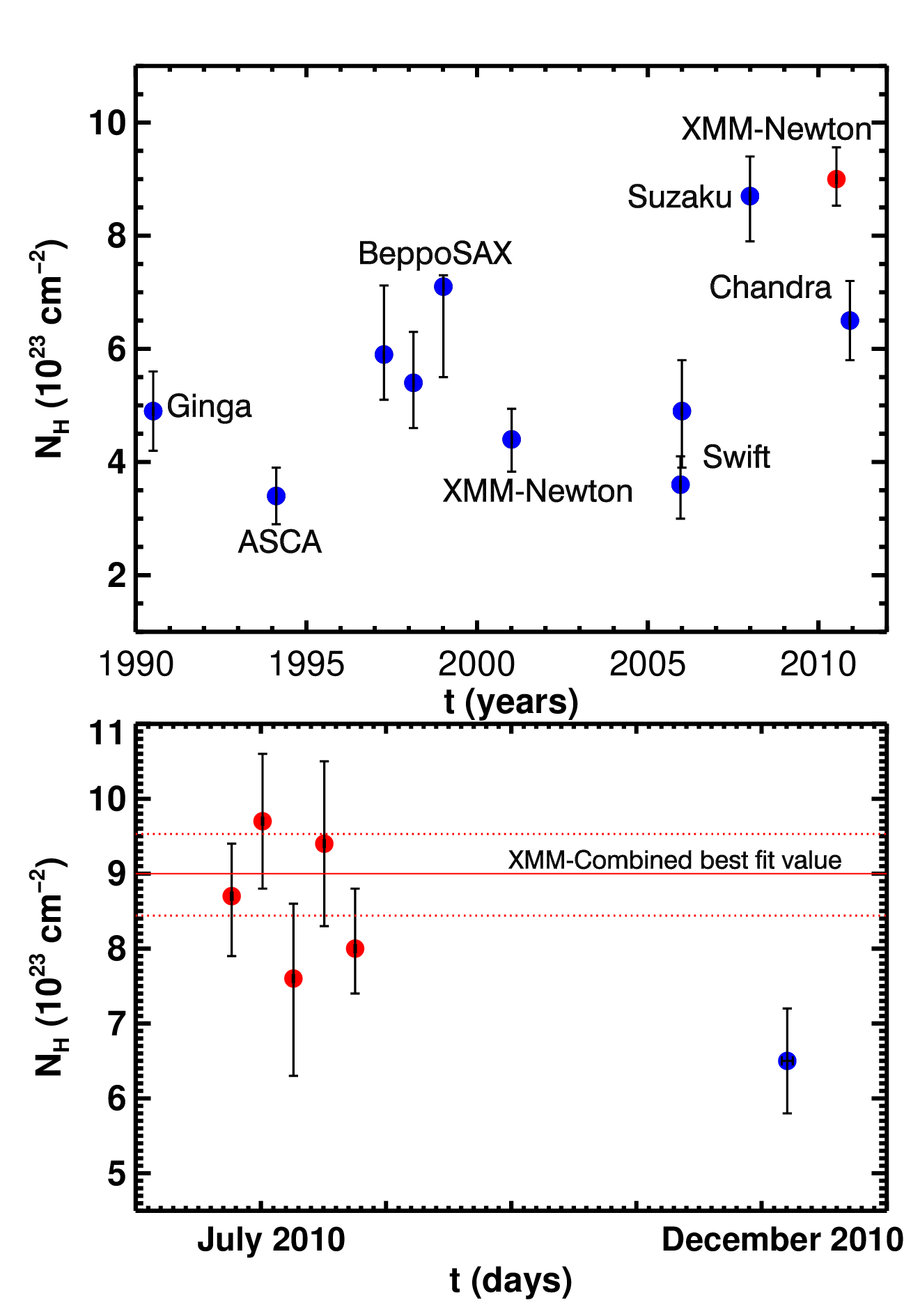,width=\columnwidth}
\caption{\label{nhplot} Column density light curves. (a) From 1990 until now the source has presented several $N_{\rm H}$ variations. Values are taken from \citet{risa02b}. A clear changing-look on time scales of years has been already discussed in B12, while for the first time a column density variation has been observed on time scales of months. (b) Column density light curve from June 2010 to December 2010. The 5 XMM-Newton observations do not show any evidence of variation in N$_{\rm H}$, while the Chandra value clearly differs from the XMM combined best fit value, which is plotted as a red solid line with errors as red dashed lines.  }
\end{center}
\end{figure}

\section{Physical discussion}
From the X-ray data analysis presented above a column density variation in a time scale of months is evident. We are going to describe, in the following, the physical implications of this result. Changes in absorbing column density are due to two different physical processes: a variation of the ionizing primary radiation, which causes the variation in the ionization state of the absorber or variations in the amount of absorbing gas along the line of sight. In the case of NGC 4507 the first physical scenario can be clearly ruled out because the difference in the 2-10 keV fluxes between the observations is not significant enough to justify a variation in the primary ionizing radiation.

\begin{figure}
\begin{center}
\epsfig{file=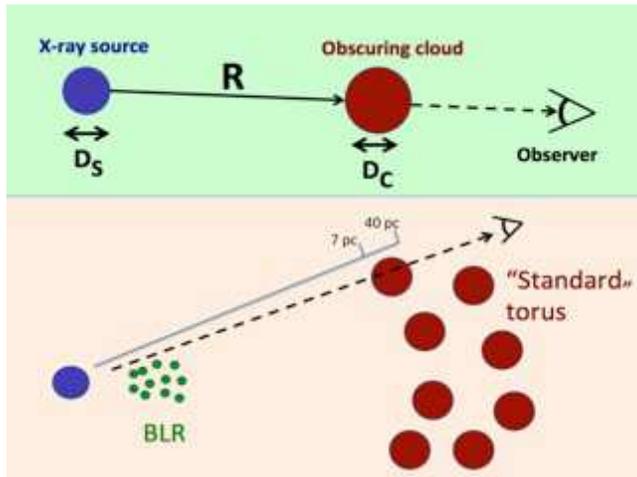,width=\columnwidth}
\caption{\label{structure} Schematic view of the circumnuclear absorbing structure.  }
\end{center}
\end{figure}

The black hole mass of NGC 4507 is estimated by means of stellar velocity dispersion to be $4.5\times 10^7$M$_{\odot}$ \citep{mari12a}. We assume the dimensions of the X-ray emitting source D$_S$ to be 10 R$_G$. This is in agreement with continuum variability studies and disk-corona emission models, all suggesting a compact central X-ray source, confined within a few R$_G$ from the central black hole.  We also assume that the size of the obscuring cloud $D_C$$\sim$$D_S$. A schematic view of the geometrical structure is shown in Fig. \ref{structure}. The transverse velocity $v_K$ for one obscuring cloud is then simply given by the linear dimension of the X-ray source, $D_S$, divided by the crossing time $T_{cr}$:\\
\begin{eqnarray}
v_K=\frac{D_C}{T_{cr}}\simeq\frac{10\ GM_{bh}}{c^2\ T_{cr}}\simeq 70\ {\rm km\ s^{-1}} M_{7.65} T^{-1}_{4},
\end{eqnarray}
where we introduced the adimensional parameters $M_{7.65}=M_{bh}/10^{7.65}M_{\odot}$ and $T_{4}\simeq1\times10^7$s = 4 months.\\
If we then consider the absorbing material located at a distance $R$ from the central X-ray source, moving with Keplerian velocity, we can calculate $R$ with the simple formula:
\begin{eqnarray}
R=\frac{GM_{bh}}{v^2_K}=\frac{GM_{bh} T^2_{cr}}{10^2 R^2_G}\simeq40\ {\rm pc} \ M_{7.65} R_{10}^{-2} T^{2}_{4},
\end{eqnarray}
where $R_{10}=D_{S}/10\ R_G$. 
In the case of NGC 4507 the lower limit on the crossing time is 1.5 months, the time interval between the first and last observation of our XMM-Newton campaign, during which the column density is maximal and nearly constant.  The upper limit to the uncovering time ($\sim D_S/v_K$) is 4 months, time interval between the fifth XMM-Newton observation (August 2011) and the Chandra one (December 2011).\\
Using these limits in the relations above we get a lower limit of $R=7\ pc \ M_{7.65} R_{10}^{-2}$ and an upper limit of $R=40\ pc \ M_{7.65} R_{10}^{-2}$. These distances imply that the obscuring material is located well outside the BLR. To be consistent with the BLR location the obscuring cloud should have a linear size $D_C> 20 D_S$. If we assume a typical BLR density $n_{\rm e} \sim 10^9$ cm$^{-3}$ \citep{oster89} and a N$_{\rm H}=2.5\times 10^{23}$cm$^{-2}$ (difference between XMM-Newton and Chandra spectra) we get a linear size for the obscuring cloud $D_C<3.5\ D_S$: these two values are clearly inconsistent. The Suzaku observation is the only one that really probes the typical BLR timescales in NGC 4507, but with an average column density of $\sim10^{24}$ cm$^{-2}$ it is not possible to disentangle any inner component with N$_H\sim10^{22}-10^{23}$ cm$^{-2}$.\\
We reduced and analyzed two 10ks long Swift-XRT observations on the 24$^{\rm th}$ and 30$^{\rm th}$ of December 2005  ({\sc{obsid} } 00035465001 and 00035465003 respectively), to check for column density variations on timescales of days. The baseline model we used to reproduce the data is a simple absorbed power law, a Gaussian emission line and a further soft power law for spectral features below $\sim$3 keV. Data quality does not allow for a clear investigation of the $\Gamma$-N$_{\rm H}$ parameter space, so we adopted a fixed value of $\Gamma=1.8$. Only a marginal column density variation (at 1$\sigma$ confidence level) is found between the two Swift observations and the two measurements of N$_{\rm H}$ are consistent at the 90\% confidence level (Fig. \ref{nhplot}). 
The lack of significant variability on short time scales does not imply that a BLR component does not exist at all, it suggests indeed a much more complex environment of absorbing structures, as already discussed in \citet{narris11} for the dwarf Seyfert galaxy NGC 4395. \\
The obscuring clouds' velocities we measured (70-170 km/s $M_{7.65}$) are at least one order of magnitude smaller than the typical BLR clouds' velocities. The absorption variability is due to circumnuclear material which is located at distances consistent with the putative torus \citep{antonucci93}.  Such material cannot be located at much larger scales (i.e. dust lanes) since the NLR is not significantly affected by reddening, as inferred by the observed H$\alpha$/H$\beta$ ratio \citep{kehe01}.\\
Since the presence of a Compton-thick material around the nucleus is invariably accompanied by a neutral iron narrow K$\alpha$-emission line and a cold reflection emission it is interesting to point out the fact that the spatial scale of the obscuring material in NGC 4507 is consistent with the distance of the reflector observed with {\it Chandra} in the nearby Compton-thick Sy2 NGC 4945 \citep{mari12b}.\\
The existence of a more complex structure surrounding the central engine of AGN rather than the one predicted by the unification model has been proposed and widely discussed in the past few years \citep{maiorieke95, elvis00, matt00b} and recently in \citet{bianchi07b,risael10a,bmr12, elvis12}. \\
Accordingly to these models, only absorbing matter on very different scales can be responsible for the wide phenomenology of column density variations, leading to an overall scenario where different absorbers/reflectors are responsible for the spectral changes in the Seyfert 2 galaxies observed so far. \\
The column density variation we measured in NGC 4507 is a further piece that can be added to the puzzle. In our analysis the lack of any variation on short time scales (hours, days) excludes an absorber located in the BLR while the change on time scales of months leads to an absorber much farther from the X-ray source, differently with respect to other variable objects \citep[e.g. NGC 1365, NGC 4151, UGC 4203:][]{ris05, ris07, ris09,puc07,risael10} on long time scales (months, years) and rapidly changing on short ones observed so far. Our analysis provides further evidence that a universal circumnuclear structure of absorbing matter is therefore not suited for taking into account all the observed phenomenology on absorption variability in AGN. While it is true that absorption must occur on different scales, not all the objects present evidence of absorption from all the possible scales. \\

\section{Conclusions}
We reported in this paper the analysis of 5 XMM-Newton observations spanning a period of 6 weeks and a Chandra observation performed 4 months afterwards of the obscured AGN in NGC 4507. This source had shown strong column density variations in time scales of years (from 1990 until 2007) and none in shorter time scales during the 3 days of Suzaku monitoring. We therefore investigated time scales ranging from 1.5 up to 4 months, looking for absorbing structures located farther from the innermost X-ray source.

Our results can be summarized as follows:
\begin{itemize}
\item{ a column density variation of $|\Delta N_{\rm H}|= 2.5\times10^{23}$ cm$^{-2}$ at a 3$\sigma$ confidence level on a time interval between 1.5 and 4 months has been measured. Such time scales lead to distances of the absorber ranging from $R=(7-40)\ M_{7.65}R^{-2}_{10}\ pc$ with corresponding velocities of $70-170$ km/s $M_{7.65}$. These distances imply that the obscuring material is located well outside the BLR, and suggest a much more complex environment of absorbing structures;} 
\item {the distances we inferred suggest that a single, universal structure of the absorber is not enough to reproduce the X-ray absorption variability of this AGN. Different reflectors/absorbers are responsible for the observed X-ray features.}
\end{itemize}
In the next future, following the results presented in \citet{risa02}, a monitoring of all the sources that have shown changes on time scales of years but none on shorter (hours-days) can be performed. A broad band, time-resolved study of AGN is fundamental for a better understanding of the complex circumnuclear material and its interaction and response to the primary radiation.
\section*{ACKNOWLEDGEMENTS}
The authors thank both the referees for their comments, that greatly improved this paper. This work was partially supported by NASA grants NNX11AC85G and GO2-13124X.
\bibliographystyle{mn2e}
\bibliography{sbs} 

\begin{thebibliography}{}

\bibitem[\protect\citeauthoryear{{Anders} \& {Grevesse}}{{Anders} \&
  {Grevesse}}{1989}]{angr89}
{Anders} E.,  {Grevesse} N.,  1989, \gca, 53, 197

\bibitem[\protect\citeauthoryear{{Antonucci}}{{Antonucci}}{1993}]{antonucci93}
{Antonucci} R.,  1993, \araa, 31, 473

\bibitem[\protect\citeauthoryear{{Arnaud}}{{Arnaud}}{1996}]{xspec}
{Arnaud} K.~A.,  1996, in ASP Conf. Ser. 101: Astronomical Data Analysis
  Software and Systems V {XSPEC: The First Ten Years}.
p.~17

\bibitem[\protect\citeauthoryear{{Awaki}, {Kunieda}, {Tawara} \&
  {Koyama}}{{Awaki} et~al.}{1991}]{awaki91b}
{Awaki} H.,  {Kunieda} H.,  {Tawara} Y.,    {Koyama} K.,  1991, \pasj, 43, L37

\bibitem[\protect\citeauthoryear{{Bianchi}, {Chiaberge}, {Evans}, {Guainazzi},
  {Baldi}, {Matt} \& {Piconcelli}}{{Bianchi} et~al.}{2010}]{bianchi10}
{Bianchi} S.,  {Chiaberge} M.,  {Evans} D.~A.,  {Guainazzi} M.,  {Baldi} R.~D.,
   {Matt} G.,    {Piconcelli} E.,  2010, \mnras, pp 418--+

\bibitem[\protect\citeauthoryear{{Bianchi}, {Chiaberge}, {Piconcelli} \&
  {Guainazzi}}{{Bianchi} et~al.}{2007}]{bianchi07b}
{Bianchi} S.,  {Chiaberge} M.,  {Piconcelli} E.,    {Guainazzi} M.,  2007,
  \mnras, 374, 697

\bibitem[\protect\citeauthoryear{{Bianchi} \& {Guainazzi}}{{Bianchi} \&
  {Guainazzi}}{2007}]{bg07}
{Bianchi} S.,  {Guainazzi} M.,  2007, in {di Salvo} T.,  {Israel} G.~L.,
  {Piersant} L.,  {Burderi} L.,  {Matt} G.,  {Tornambe} A.,   {Menna} M.~T.,
  eds, The Multicolored Landscape of Compact Objects and Their Explosive
  Origins Vol.~924 of American Institute of Physics Conference Series, {The
  nature of the soft X-ray emission in obscured AGN}.
pp 822--829

\bibitem[\protect\citeauthoryear{{Bianchi}, {Guainazzi} \&
  {Chiaberge}}{{Bianchi} et~al.}{2006}]{bianchi06}
{Bianchi} S.,  {Guainazzi} M.,    {Chiaberge} M.,  2006, \aap, 448, 499

\bibitem[\protect\citeauthoryear{{Bianchi}, {Maiolino} \& {Risaliti}}{{Bianchi}
  et~al.}{2012}]{bmr12}
{Bianchi} S.,  {Maiolino} R.,    {Risaliti} G.,  2012, Advances in Astronomy,
  2012

\bibitem[\protect\citeauthoryear{{Bianchi}, {Piconcelli}, {Chiaberge},
  {Bail{\'o}n}, {Matt} \& {Fiore}}{{Bianchi} et~al.}{2009}]{bianchi09c}
{Bianchi} S.,  {Piconcelli} E.,  {Chiaberge} M.,  {Bail{\'o}n} E.~J.,  {Matt}
  G.,    {Fiore} F.,  2009, \apj, 695, 781

\bibitem[\protect\citeauthoryear{{Braito}, {Ballo}, {Reeves}, {Ptak},
  {Risaliti} \& {Turner}}{{Braito} et~al.}{2012}]{braito12}
{Braito} V.,  {Ballo} L.,  {Reeves} J.~N.,  {Ptak} A.,  {Risaliti} G.,
  {Turner} T.~J.,  2012, ArXiv e-prints

\bibitem[\protect\citeauthoryear{{Comastri}, {Vignali}, {Cappi}, {Matt},
  {Audano}, {Awaki} \& {Ueno}}{{Comastri} et~al.}{1998}]{comastri98}
{Comastri} A.,  {Vignali} C.,  {Cappi} M.,  {Matt} G.,  {Audano} R.,  {Awaki}
  H.,    {Ueno} S.,  1998, \mnras, 295, 443

\bibitem[\protect\citeauthoryear{{Dickey} \& {Lockman}}{{Dickey} \&
  {Lockman}}{1990}]{dl90}
{Dickey} J.~M.,  {Lockman} F.~J.,  1990, \araa, 28, 215

\bibitem[\protect\citeauthoryear{{Elvis}}{{Elvis}}{2000}]{elvis00}
{Elvis} M.,  2000, \apj, 545, 63

\bibitem[\protect\citeauthoryear{{Elvis}}{{Elvis}}{2012}]{elvis12}
{Elvis} M.,  2012, in {Chartas} G.,  {Hamann} F.,   {Leighly} K.~M.,  eds, AGN
  Winds in Charleston Vol.~460 of Astronomical Society of the Pacific
  Conference Series, {Quasar Structure Emerges from the Three Forms of
  Radiation Pressure}.
p.~186

\bibitem[\protect\citeauthoryear{{Elvis}, {Risaliti}, {Nicastro}, {Miller},
  {Fiore} \& {Puccetti}}{{Elvis} et~al.}{2004}]{elvis04}
{Elvis} M.,  {Risaliti} G.,  {Nicastro} F.,  {Miller} J.~M.,  {Fiore} F.,
  {Puccetti} S.,  2004, \apjl, 615, L25

\bibitem[\protect\citeauthoryear{{Fruscione}, {McDowell}, {Allen},
  {Brickhouse}, {Burke}, {Davis}, {Durham}, {Elvis}, {Galle}, {Harris},
  {Huenemoerder}, {Houck}, {Ishibashi}, {Karovska}, {Nicastro}, {Noble},
  {Nowak} \& {Primini}}{{Fruscione} et~al.}{2006}]{ciao}
{Fruscione} A.,  {McDowell} J.~C.,  {Allen} G.~E.,  {Brickhouse} N.~S.,
  {Burke} D.~J.,  {Davis} J.~E.,  {Durham} N.,  {Elvis} M.,  {Galle} E.~C.,
  {Harris} D.~E.,  {Huenemoerder} D.~P.,  {Houck} J.~C.,  {Ishibashi} B.,
  {Karovska} M.,  {Nicastro} F.,  {Noble} M.~S.,  {Nowak} M.~A.,    {Primini}
  F.~A.,  2006, in Observatory Operations: Strategies, Processes, and Systems.
  Edited by Silva, David R.; Doxsey, Rodger E.. Proceedings of the SPIE, Volume
  6270, pp. 62701V (2006). Vol.~6270 of Presented at the Society of
  Photo-Optical Instrumentation Engineers (SPIE) Conference, {CIAO: Chandra's
  data analysis system}

\bibitem[\protect\citeauthoryear{{Gabriel}, {Denby}, {Fyfe}, {Hoar}, {Ibarra},
  {Ojero}, {Osborne}, {Saxton}, {Lammers} \& {Vacanti}}{{Gabriel}
  et~al.}{2004}]{gabr04}
{Gabriel} C.,  {Denby} M.,  {Fyfe} D.~J.,  {Hoar} J.,  {Ibarra} A.,  {Ojero}
  E.,  {Osborne} J.,  {Saxton} R.~D.,  {Lammers} U.,    {Vacanti} G.,  2004, in
  {F.~Ochsenbein, M.~G.~Allen, \& D.~Egret} ed., Astronomical Data Analysis
  Software and Systems (ADASS) XIII Vol.~314 of Astronomical Society of the
  Pacific Conference Series, {The XMM-Newton SAS - Distributed Development and
  Maintenance of a Large Science Analysis System: A Critical Analysis}.
pp 759--+

\bibitem[\protect\citeauthoryear{{Garmire}, {Bautz}, {Ford}, {Nousek} \&
  {Ricker}}{{Garmire} et~al.}{2003}]{acis}
{Garmire} G.~P.,  {Bautz} M.~W.,  {Ford} P.~G.,  {Nousek} J.~A.,    {Ricker}
  G.~R.,  2003, in X-Ray and Gamma-Ray Telescopes and Instruments for
  Astronomy. Edited by Joachim E. Truemper, Harvey D. Tananbaum. Proceedings of
  the SPIE, Volume 4851, p. 28-44 {Advanced CCD imaging spectrometer (ACIS)
  instrument on the Chandra X-ray Observatory}

\bibitem[\protect\citeauthoryear{{Guainazzi} \& {Bianchi}}{{Guainazzi} \&
  {Bianchi}}{2007}]{guabia07}
{Guainazzi} M.,  {Bianchi} S.,  2007, \mnras, 374, 1290

\bibitem[\protect\citeauthoryear{{Kewley}, {Heisler}, {Dopita} \&
  {Lumsden}}{{Kewley} et~al.}{2001}]{kehe01}
{Kewley} L.~J.,  {Heisler} C.~A.,  {Dopita} M.~A.,    {Lumsden} S.,  2001,
  \apjs, 132, 37

\bibitem[\protect\citeauthoryear{{Kriss}, {Canizares} \& {Ricker}}{{Kriss}
  et~al.}{1980}]{kriss80}
{Kriss} G.~A.,  {Canizares} C.~R.,    {Ricker} G.~R.,  1980, \apj, 242, 492

\bibitem[\protect\citeauthoryear{{Krolik} \& {Begelman}}{{Krolik} \&
  {Begelman}}{1988}]{krbe88}
{Krolik} J.~H.,  {Begelman} M.~C.,  1988, \apj, 329, 702

\bibitem[\protect\citeauthoryear{{Magdziarz} \& {Zdziarski}}{{Magdziarz} \&
  {Zdziarski}}{1995}]{mz95}
{Magdziarz} P.,  {Zdziarski} A.~A.,  1995, \mnras, 273, 837

\bibitem[\protect\citeauthoryear{{Maiolino} \& {Rieke}}{{Maiolino} \&
  {Rieke}}{1995}]{maiorieke95}
{Maiolino} R.,  {Rieke} G.~H.,  1995, \apj, 454, 95

\bibitem[\protect\citeauthoryear{{Marinucci}, {Bianchi}, {Matt}, {Fabian},
  {Iwasawa}, {Miniutti} \& {Piconcelli}}{{Marinucci}
  et~al.}{2011}]{marinucci11}
{Marinucci} A.,  {Bianchi} S.,  {Matt} G.,  {Fabian} A.~C.,  {Iwasawa} K.,
  {Miniutti} G.,    {Piconcelli} E.,  2011, \aap, 526, A36+

\bibitem[\protect\citeauthoryear{{Marinucci}, {Bianchi}, {Nicastro}, {Matt} \&
  {Goulding}}{{Marinucci} et~al.}{2012}]{mari12a}
{Marinucci} A.,  {Bianchi} S.,  {Nicastro} F.,  {Matt} G.,    {Goulding} A.~D.,
   2012, \apj, 748, 130

\bibitem[\protect\citeauthoryear{{Marinucci}, {Risaliti}, {Wang}, {Nardini},
  {Elvis}, {Fabbiano}, {Bianchi} \& {Matt}}{{Marinucci} et~al.}{2012}]{mari12b}
{Marinucci} A.,  {Risaliti} G.,  {Wang} J.,  {Nardini} E.,  {Elvis} M.,
  {Fabbiano} G.,  {Bianchi} S.,    {Matt} G.,  2012, \mnras, 423, L6

\bibitem[\protect\citeauthoryear{{Matt}}{{Matt}}{2000}]{matt00b}
{Matt} G.,  2000, \aap, 355, L31

\bibitem[\protect\citeauthoryear{{Matt}}{{Matt}}{2002}]{matt02}
{Matt} G.,  2002, \mnras, 337, 147

\bibitem[\protect\citeauthoryear{{Matt}, {Bianchi}, {D'Ammando} \&
  {Martocchia}}{{Matt} et~al.}{2004}]{matt04b}
{Matt} G.,  {Bianchi} S.,  {D'Ammando} F.,    {Martocchia} A.,  2004, \aap,
  421, 473

\bibitem[\protect\citeauthoryear{{Nardini} \& {Risaliti}}{{Nardini} \&
  {Risaliti}}{2011}]{narris11}
{Nardini} E.,  {Risaliti} G.,  2011, \mnras, 417, 2571

\bibitem[\protect\citeauthoryear{{Osterbrock}}{{Osterbrock}}{1989}]{oster89}
{Osterbrock} D.~E.,  1989, {Astrophysics of gaseous nebulae and active galactic
  nuclei}

\bibitem[\protect\citeauthoryear{{Piconcelli}, {Jimenez-Bail{\' o}n},
  {Guainazzi}, {Schartel}, {Rodr{\'{\i}}guez-Pascual} \& {Santos-Lle{\'
  o}}}{{Piconcelli} et~al.}{2004}]{pico04}
{Piconcelli} E.,  {Jimenez-Bail{\' o}n} E.,  {Guainazzi} M.,  {Schartel} N.,
  {Rodr{\'{\i}}guez-Pascual} P.~M.,    {Santos-Lle{\' o}} M.,  2004, \mnras,
  351, 161

\bibitem[\protect\citeauthoryear{{Protassov}, {van Dyk}, {Connors}, {Kashyap}
  \& {Siemiginowska}}{{Protassov} et~al.}{2002}]{prota02}
{Protassov} R.,  {van Dyk} D.~A.,  {Connors} A.,  {Kashyap} V.~L.,
  {Siemiginowska} A.,  2002, \apj, 571, 545

\bibitem[\protect\citeauthoryear{{Puccetti}, {Fiore}, {Risaliti}, {Capalbi},
  {Elvis} \& {Nicastro}}{{Puccetti} et~al.}{2007}]{puc07}
{Puccetti} S.,  {Fiore} F.,  {Risaliti} G.,  {Capalbi} M.,  {Elvis} M.,
  {Nicastro} F.,  2007, \mnras, 377, 607

\bibitem[\protect\citeauthoryear{{Risaliti}}{{Risaliti}}{2002}]{risa02}
{Risaliti} G.,  2002, \aap, 386, 379

\bibitem[\protect\citeauthoryear{{Risaliti} \& {Elvis}}{{Risaliti} \&
  {Elvis}}{2010}]{risael10a}
{Risaliti} G.,  {Elvis} M.,  2010, \aap, 516, A89

\bibitem[\protect\citeauthoryear{{Risaliti}, {Elvis}, {Bianchi} \&
  {Matt}}{{Risaliti} et~al.}{2010}]{risael10}
{Risaliti} G.,  {Elvis} M.,  {Bianchi} S.,    {Matt} G.,  2010, \mnras, 406,
  L20

\bibitem[\protect\citeauthoryear{{Risaliti}, {Elvis}, {Fabbiano}, {Baldi} \&
  {Zezas}}{{Risaliti} et~al.}{2005}]{ris05}
{Risaliti} G.,  {Elvis} M.,  {Fabbiano} G.,  {Baldi} A.,    {Zezas} A.,  2005,
  \apjl, 623, L93

\bibitem[\protect\citeauthoryear{{Risaliti}, {Elvis}, {Fabbiano}, {Baldi},
  {Zezas} \& {Salvati}}{{Risaliti} et~al.}{2007}]{ris07}
{Risaliti} G.,  {Elvis} M.,  {Fabbiano} G.,  {Baldi} A.,  {Zezas} A.,
  {Salvati} M.,  2007, \apjl, 659, L111

\bibitem[\protect\citeauthoryear{{Risaliti}, {Elvis} \& {Nicastro}}{{Risaliti}
  et~al.}{2002}]{risa02b}
{Risaliti} G.,  {Elvis} M.,    {Nicastro} F.,  2002, \apj, 571, 234

\bibitem[\protect\citeauthoryear{{Risaliti}, {Miniutti}, {Elvis}, {Fabbiano},
  {Salvati}, {Baldi}, {Braito}, {Bianchi}, {Matt}, {Reeves}, {Soria} \&
  {Zezas}}{{Risaliti} et~al.}{2009}]{ris09}
{Risaliti} G.,  {Miniutti} G.,  {Elvis} M.,  {Fabbiano} G.,  {Salvati} M.,
  {Baldi} A.,  {Braito} V.,  {Bianchi} S.,  {Matt} G.,  {Reeves} J.,  {Soria}
  R.,    {Zezas} A.,  2009, \apj, 696, 160

\bibitem[\protect\citeauthoryear{{Smith}, {Brickhouse}, {Liedahl} \&
  {Raymond}}{{Smith} et~al.}{2001}]{apec}
{Smith} R.~K.,  {Brickhouse} N.~S.,  {Liedahl} D.~A.,    {Raymond} J.~C.,
  2001, \apjl, 556, L91

\bibitem[\protect\citeauthoryear{{Str{\"u}der}, {Briel}, {Dennerl}, {Hartmann},
  {Kendziorra}, {Meidinger}, {Pfeffermann}, {Reppin}, {Aschenbach},
  {Bornemann}, {Br{\" a}uninger}, {Burkert} \& {Elender}}{{Str{\"u}der}
  et~al.}{2001}]{struder01}
{Str{\"u}der} L.,  {Briel} U.,  {Dennerl} K.,  {Hartmann} R.,  {Kendziorra} E.,
   {Meidinger} N.,  {Pfeffermann} E.,  {Reppin} C.,  {Aschenbach} B.,
  {Bornemann} W.,  {Br{\" a}uninger} H.,  {Burkert} W.,    {Elender} M.,  2001,
  \aap, 365, L18

\bibitem[\protect\citeauthoryear{{Turner}, {Abbey}, {Arnaud}, {Balasini},
  {Barbera}, {Belsole}, {Bennie}, {Bernard}, {Bignami}, {Boer}, {Briel},
  {Butler}, {Cara}, {Chabaud}, {Cole}, {Collura}, {Conte}, {Cros} \&
  {Denby}}{{Turner} et~al.}{2001}]{turner01}
{Turner} M.~J.~L.,  {Abbey} A.,  {Arnaud} M.,  {Balasini} M.,  {Barbera} M.,
  {Belsole} E.,  {Bennie} P.~J.,  {Bernard} J.~P.,  {Bignami} G.~F.,  {Boer}
  M.,  {Briel} U.,  {Butler} I.,  {Cara} C.,  {Chabaud} C.,  {Cole} R.,
  {Collura} A.,  {Conte} M.,  {Cros} A.,    {Denby} 2001, \aap, 365, L27

\bibitem[\protect\citeauthoryear{{Turner}, {George}, {Nandra} \&
  {Mushotzky}}{{Turner} et~al.}{1997}]{turner97}
{Turner} T.~J.,  {George} I.~M.,  {Nandra} K.,    {Mushotzky} R.~F.,  1997,
  \apjs, 113, 23

\end{thebibliography}

\end{document}